\documentclass[twocolumn]{aa}
\usepackage{epsfig}
\def\roxa{ROXA~J081009.9+384757.0}
\def\sdssstar{SDSS~J081010.9+384745.6}
\def\sdss{SDSS~J081009.9+384757.0}
\def\rass{RX~J0810.2+3847}
\def\first{FIRST~J081009.9+384756}

\newcommand{\lsim}{{\lower.5ex\hbox{$\; \buildrel < \over \sim \;$}}}
\newcommand{\gsim}{{\lower.5ex\hbox{$\; \buildrel > \over \sim \;$}}}
\def\sw{Swift~}

\def\sax{{\it Beppo}SAX~}
\def\aro{{$\alpha_{\rm ro}$~}}

\def\arx{{$\alpha_{\rm rx}$~}}
\def\erg{{erg~s$^{-1}$~}}
\def\ksm{{km~s$^{-1}$~Mpc$^{-1}$}}
\def\ergs{{erg~cm$^{-2}$s$^{-1}$~}}
\def\ergsjy{{erg~cm$^{-2}$s$^{-1}$Jy$^{-1}$~}}

\def\fxfr{$f_{\rm x}/f_{\rm r}$}
\def\fxfmu{$f_{\rm x}/f_{94GHz}$~}
\begin{document}

\title{\roxa: a $10^{47}$ \erg blazar with hard X-ray synchrotron peak or a new type of radio
loud AGN?}
\author{P.~Giommi\inst{1}
 \and E.~Massaro\inst{2,1}
 \and P.~Padovani\inst{3}
 \and M.~Perri\inst{1}
 \and E.~Cavazzuti\inst{1}
 \and S.~Turriziani\inst{4}
 \and G.~Tosti\inst{5}
 \and S.~Colafrancesco \inst{6}
 \and G.~Tagliaferri\inst{7}
 \and G.~Chincarini\inst{7}
 \and D.N.~Burrows\inst{8}
 \and M.~McMath Chester\inst{8}
 \and N.~Gehrels\inst{9}
 }
 \institute{ ASI Science Data Center, ESRIN, I-00044 Frascati, Italy
 \and       Dipartimento di Fisica, Universit\`a "La Sapienza",
               P.le A. Moro 2, I-00185 Roma, Italy.
 \and       European Southern Observatory, Karl Schwarzschild-Str. 2, Garching bei M\"unchen, Germany
 \and       Universit\`a degli studi di Roma, Tor Vergata, Dip. Fisica,
             via della ricerca scientifica 1,  00133  Roma, Italy.
 \and       Dipartimento di Fisica, Universit\'a di Perugia, Via A. Pascoli,  Perugia, Italy.
 \and      INAF - Osservatorio Astronomico di Roma, via Frascati 33, I-00040 Monteporzio, Italy
 \and       INAF, Osservatorio Astronomico di Brera, via Bianchi 46, 23807 Merate, Italy
 \and       Department of Astronomy and Astrophysics,  Pennsylvania  State
              University, USA.
 \and      NASA/Goddard Space Flight Center, Greenbelt, Maryland 20771, USA.
  }

\offprints{paolo.giommi@asi.it}
\date{Received; accepted}

\markboth{P.Giommi: \roxa: ~ the brightest High Energy Peaked Blazar }
{ P.Giommi: \roxa: ~ the brightest synchrotron emitter}

\abstract {We report the discovery of \roxa= \sdss, a z=3.95 blazar with a highly unusual
Spectral Energy Distribution (SED). This object was first noticed as a probable high
\fxfr, high-luminosity blazar within the error region of a $\approx 10^{-12}$ \ergs ROSAT
source which, however, also included a much brighter late-type star. We describe the
results of a recent \sw observation that establishes beyond doubt that the correct
counterpart of the X-ray source is the flat spectrum radio quasar. With a luminosity well
in excess of  $10^{47}$ \erg \roxa~ is therefore one of the most luminous blazars known.
We consider various possibilities for the nature of the electromagnetic emission from
this source. In particular, we show that the SED is consistent with that of a blazar with
synchrotron power peaking in the hard X-ray band. If this is indeed the case, the
combination of high-luminosity and synchrotron peak in the hard-X-ray band contradicts
the claimed anti-correlation between luminosity and position of the synchrotron peak
usually referred to as the "blazar sequence". An alternative possibility is that the
X-rays are not due to synchrotron emission, in this case the very peculiar SED of \roxa
~would make it the first example of a new class of radio loud AGN.

}
\authorrunning{P. Giommi et al.}
\titlerunning{\roxa: ~ the brightest High Energy Peaked Blazar
}


\maketitle

\section{Introduction}

Active Galactic Nuclei (AGN) come in two main categories depending on the type of
radiation we observe from them. AGN in which the observed power is dominated by thermal
radiation generated in a disk of material accreted onto a super-massive black-hole (or
Thermal-Emission Dominated AGN, the large majority of the population), and AGN in which
the observed power is instead dominated by non-thermal radiation emitted by highly
energetic particles in a jet of material that moves away from the central black hole at
relativistic speeds (or Non-Thermal-Emission Dominated AGN, \cite{GioCol06}). Within this
paradigm, blazars are the small subset of Non-Thermal-Emission Dominated AGN that happen
to be viewed at a small angle with respect to the jet axis. Because of this very special
situation, the observed radiation from blazars is strongly affected by relativistic
amplification and time contraction that are at the root of their extreme properties, such
as super-luminal motion, strong rapid variability and high polarization
(\cite{bla78,Urry95}).
Blazars comprise the class of BL Lac objects, distinguished by featureless optical
spectra, and of Flat Spectrum Radio Quasars (FSRQs) which instead share all the usual
broad emission lines that are typical of AGNs.

One specific  property of blazars is that their non-thermal emission spans the entire
electromagnetic spectrum. In fact, their Spectral Energy Distribution (SED), when
represented in the $Log(\nu)-Log[\nu f(\nu)]$ space, is composed of a synchrotron
low-energy component that usually peaks between the far infrared and the X-ray band,
followed by an inverse-Compton component that has its maximum in the hard X-ray band or
at higher energies, and extends into the $\gamma$-ray or even the TeV band. BL Lacs with
synchrotron peak located at low energy are usually called Low-energy peaked BL Lacs
(LBL), while those where the synchrotron component reaches the X-ray band are called
High-energy peaked BL Lacs  (HBL \cite{P95}). In the following we extend this terminology
to all types of blazars. In fact, the SED of almost the totality of FSRQs is very similar
to that of BL Lacs of the LBL type, although a number of these quasars (also called
HFSRQs or "HBL-like"  FSRQs) have recently been found by e.g. \cite{P03}. These objects
are, however, rare and their synchrotron emission has never been found to reach the
extreme energies observed in TeV detected BL Lacs like MKN501 (e.g.,
\cite{Pian98,Mas04b}), 1H~1100-230, MKN~421 and H~1426+428 (\cite{Tramacere06}).
A recent survey specifically designed to search for extreme HBL sources resulted in the
selection of  about 150 BL Lacs and no FSRQs (\cite{Gio05}).

To explain the different SEDs observed in radio and X-ray selected samples of Blazars,
\cite{Fos98} proposed that blazars form a sequence where the peak energy of the
synchrotron component is inversely related to the observed luminosity of the object.
\cite{Ghi98} put this hypothesis in a theoretical framework where the sequence would be
the result  of a competitive balance between particle acceleration and radiative cooling.
In this view, hence, only low-power BL Lac objects can reach synchrotron peak energies in
the X-ray band, whereas the peak energy of high-luminosity FSRQs cannot go much beyond
the far infrared frequency range.

In this paper we report the results of a \sw  (\cite{Gehrels04}) observation of one of
the most luminous known FSRQ in the X-ray band, \roxa = \sdss , which has a very large
X-ray to radio flux ratio \fxfr $\simeq 5 \times 10^{-11}$ \ergsjy, typical of HBL
objects, with synchrotron peak in the X-ray band. We discuss here the impact of this
source  on the blazar sequence hypothesis.
Throughout this paper we use a flat, vacuum-dominated CDM cosmology with  H$_0$=70 \ksm,
$\Omega_{M}$=0.3, $\Omega_{\Lambda}$=0.7.

\section{\sdss ~and \rass}

\begin{figure}
 \begin{center}
 \hspace{-0.3cm}
 \epsfig{file=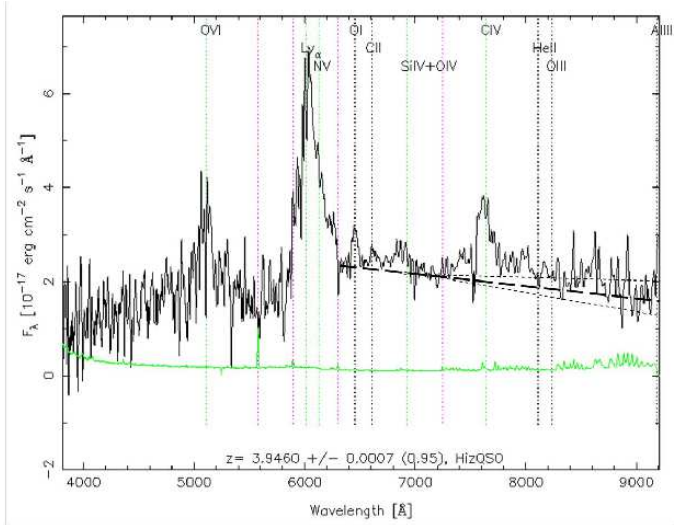,height=7.cm,angle=0}
 \end{center}
 \caption{The SDSS optical spectrum of \roxa.
The dashed line plotted to the right of the Lyman-$\alpha$ line is an estimate of the
continuum based on the composite quasar spectrum template of \cite{Vandenberk01}. The
dotted lines represent conservative limits to the slope of the continuum. See text for
details. }
 \label{sdss_spectrum}
\end{figure}

\sdss~ is a high redshift (z=3.95) object coincident with the compact radio source \first
 ~which has a radio flux of  27-30 mJy  at 1.4 GHz (from the FIRST and NVSS surveys,
\cite{White97,Con98}) and 22 mJy at 5GHz (from the GB6 catalog, \cite{Gre96}).
Fig.\ref{sdss_spectrum} shows its optical spectrum (taken from the Sloan Digital Sky
Survey (SDSS) archive, Data Release 3 (\cite{Sloan_dr3}) which clearly displays Lyman
$\alpha $ forest absorption features together with the strong Hydrogen, Oxygen  and
Carbon emission lines that are typical of distant QSOs. Since the radio spectral index of
\first ~is flat ($\alpha_{1.4-5.GHz}\simeq 0.2$, $f(\nu) \propto \nu^{-\alpha} $) and the
radio to optical spectral index \aro = 0.48, \sdss~ can safely be classified as a FSRQ.
\cite{Proc05} report that \sdss~ is a damped Lyman-$\alpha $ system with absorbing clouds
of material located at redshifts z=1.69, 3.0, 3.2 and 3.9. We estimate the slope of the
optical-UV continuum to be  $\alpha_o \simeq 0.5$, (represented by  the dashed line in
Fig. \ref{sdss_spectrum}) assuming that the minimum flux between the CIV and SiIV lines
at $\sim 7200$ \AA, ($\sim 1450$ \AA~ in the rest frame of the QSO), is the continuum
emission as in the composite QSO spectrum built by  \cite{Vandenberk01}. Given the
presence of strong emission lines it is not simple to associate an error to this slope.
The dotted lines in Fig. \ref{sdss_spectrum}, corresponding to  $\alpha_o =1.8$ and
$\alpha_o = 0.2$, represent conservative limits to the slope of the continuum that we
will use later in the paper.

The position of \sdss~ is within the error circle of \rass, one of the sources of the
ROSAT All Sky Survey  (RASS, \cite{Brinkmann00}) and is included in a sample of blazar
candidates  selected on the basis of their radio, optical and X-ray emission (ROXA, see
\cite{turr07}).
The association of the $r=19.7$ magnitude (SDSS) quasar with the RASS X-ray source is,
however, uncertain because the X-ray error circle also includes the brighter ($r$=17.3
magnitude, SDSS) late type star \sdssstar.

The X-ray flux of  \rass ~is $\simeq 10^{-12}~$\ergs in the 0.1-2.4 keV band.
\begin{figure}[t]
 \begin{center}
\epsfig{file=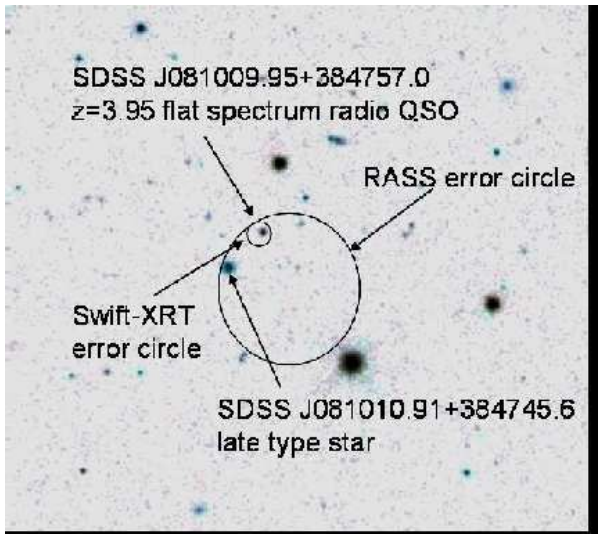,height=7.cm,angle=0}
 \end{center}
 \caption{SDSS optical image covering the 22 arc-seconds error circle of the X-ray source \rass.
 The much smaller (4 arc-seconds) \sw XRT error region associates beyond any doubt the X-ray
 source to the flat spectrum radio quasar \sdss. }
 \label{FC}
\end{figure}

\section{\sw Observations and data reduction}

The \sw  multi-frequency Gamma Ray Burst observatory (\cite{Gehrels04}) observed \rass~
as a "fill-in target" on March 3, 2006. These are short exposures carried out when the
satellite is not engaged with GRB observations. \sw carries three instruments on board:
the Burst Alert Telescope (BAT, \cite{Barthelmy05}) sensitive in the 15-150 keV band, the
X-Ray Telescope (XRT, \cite{Burrows05}) sensitive in the 0.2-10.0 keV band and capable of
measuring the position of X-ray sources with a precision of a few arc-seconds, and the UV
and Optical Telescope (UVOT, \cite{Roming05}). The purposes of the observation were: i)
to confirm that the X-ray emission is associated with the high redshift FSRQ \sdss, ii)
to measure its X-ray spectrum between 0.3 and 10 keV, and iii) to obtain simultaneous
optical--X-ray  data.

\subsection{XRT image analysis}

The \sw XRT was operated in full imaging Photon Counting mode during the entire observation
(5991 seconds).
The data were reduced using the {\it XRTDAS} software (v1.7.1) developed
at the ASI Science Data Center (ASDC) and distributed as part of the HEAsoft  6.0.4 package
by the NASA High Energy Astrophysics Archive Research Center (HEASARC).
The cleaned level-2 data has been analyzed using the XIMAGE package.

A pointlike X-ray source was clearly detected at R.A. (J2000.0)=08h 10m 10.2s Dec (J2000.0)
= +38$^{\circ}$ 47$'$ 55.3$''$ with a net count rate of  $(3.2\pm0.2)\times 10^{-2}$cts/s.
The XRT position and its associated error circle (4$''$, 90\% \cite{Moretti06}), shown in
Fig. \ref{FC} overlaid to the optical image taken from the SDSS archive, are fully
consistent with the position of \sdss~ and exclude the brighter late-type star  as the
counterpart of the X-ray source. Since \sdss~ is the only optical and radio source within
the small XRT error circle, we conclude that this quasar is the optical counterpart of
the X-ray and radio sources.

\subsection{XRT spectral analysis}

Only about 200 net counts have been detected in our short XRT observation and therefore
our spectral fits are poorly constrained.

We have selected photons with grades in the range 0-12 and used
default screening parameters to produce level 2 cleaned event files. The XRT 0.3-10 keV
spectrum was extracted from a circular region around the source with radius of 20 pixels,
which includes 90\% of the source photons.
In order to use $\chi^2$ statistics, the spectrum was rebinned to include at least 30
photons in  each energy channel. For the spectral fitting we used the  XSPEC11.3 package
and, given the very limited statistics, we only fitted the data to a simple power law
spectral model.  We first fixed the low energy absorption to the Galactic value
($N_H$=$5\times 10^{20}~$cm$^{-2}$) in the direction of the source (\cite{Dick90}). The
best fit gives a photon index $\Gamma=1.5\pm0.2$ and a reduced  $\chi^2$ = 1.0 (3
d.o.f.).

Since \sdss~ is a damped Lyman $\alpha$ system with significant amount of absorption
material located at redshifts between 1.69 and  3.9  (\cite{Proc05}), an excess of low
energy absorption compared to the Galactic value is to be expected. We then fitted the
spectrum leaving $N_H$ as a free parameter and obtained best fit values of  $N_{\rm H}$ =
$(8\pm7)\times 10^{20} ~$cm$^{-2}$ and $\Gamma=1.7\pm0.4$. Clearly, the available
statistics are not good enough to effectively constrain both spectral parameters. The
unabsorbed 2-10 keV flux is $\simeq 10^{-12}~$\ergs, similar to that observed by ROSAT,
corresponding to the very large isotropic X-ray luminosity of $1.5\times 10^{47}$ \erg.

X-ray observations of other high-redhift  quasars with ROSAT, ASCA and \sax have shown
that significant soft X-ray absorption is a common feature of radio loud quasars
(\cite{Cappi97,Fiore98,Fabian01}). This implies that the intrinsic X-ray luminosity of
\roxa ~could be higher than reported above and that its X-ray spectral slope could be
somewhat steeper than $\Gamma=1.7$. In this case also the true \fxfr~ value could be even
larger than the observed one.

\subsection{UVOT observations}

UVOT (\cite{Roming05}) is a 30 cm telescope equipped with two grisms and six broadband
filters ($V, B, U, UVW1, UVM2$, and $UVW2$).
During the \sw observation, UVOT obtained series of images in each of the available
filters.
The exposure time was 462 seconds in filter $V$ and 402 seconds in filter $B$.
The UV filters did not provide useful data since in these bands our high redshift quasar
is heavily affected by strong Lyman $\alpha$ absorption.

While all the bright optical sources visible in the SDSS image, including the late-type
star and the bright sources to the north and south of the ROSAT error circle  (see Fig.
\ref{FC}), are clearly detected in the UVOT $V$ and $B$ images, \sdss~ was not detected
even in the most sensitive $B$ filter.
Lower limits to the magnitude of \sdss ~are $V>$ 17.9 and $B >$ 20.4.
The limit in the $V$ filter is somewhat higher than those obtained with similar exposures
because the image in this case was particularly noisy.

\section{Discussion}

The source \roxa ~was discovered to be a candidate high \fxfr ~-- high luminosity blazar
in a multi-frequency survey aimed at the discovery of a large sample of new blazars by
combining data from the NVSS, RASS, GSC-2, SDSS and 2dF surveys (\cite{turr07}). The
association between the radio, optical and X-ray emission for the case of \roxa~ was
however uncertain since the error region of the X-ray source \rass~ included two
plausible optical counterparts. Our \sw observation establishes without doubt that \sdss~
is the correct counterpart of the ROSAT X-ray source and, consequently, that \roxa ~is a
high-redshift blazar with \fxfr $\simeq 5\times 10^{-11}$  \ergsjy  ($\sim 2.5\times
10^{-11}$ in the quasar's rest frame) and, consequently,  \arx $\approx $0.65. Such a
large amount of X-ray emission compared to the radio flux, even taking into account of
K-corrections, has been observed, so far, only in a very small fraction of blazars (less
than 1\% of the population, see Fig. 4 of \cite{Gio06a}, see also \cite{P03}), the vast
majority of which are BL Lacertae objects, with synchrotron emission reaching very high
energies. \roxa ~is the first high luminosity FSRQ to show such extreme broad-band
spectral characteristics.

\begin{figure}[h]
 \begin{center}
\epsfig{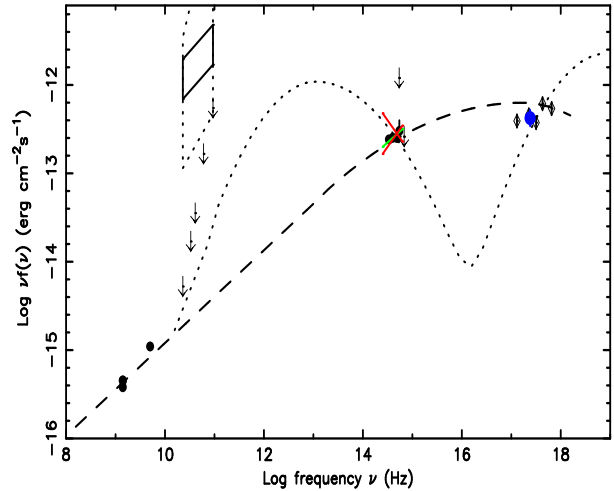}
 \end{center}
 \caption{The SED of \roxa ~built using the radio data from the NVSS, FIRST and GB6 radio surveys,
microwave upper limits (at 23, 33, 41, 61 and 94 GHz) from WMAP 3-yr data, optical
photometry from the SDSS survey, ROSAT (filled circle) and \sw XRT X-ray data (open
diamonds) and UVOT upper limits in the $V$ and $B$ filters. The boxes drawn at microwave
frequencies represent the 1 $\sigma$ and the maximum observed flux range from blazars
where the X rays are due to inverse Compton radiation (see text for details). The solid
lines in the optical band represent our best estimate and conservative upper limits to
the slope of the continuum (see also Fig. \ref{sdss_spectrum}).}
 \label{SED}
\end{figure}

We have built the overall SED of \roxa~ (see Fig. \ref{SED}) by combining the
simultaneous XRT data points and UVOT upper limits with non-simultaneous radio (FIRST,
NVSS and GB6) optical, (from SDSS magnitudes), ROSAT data and WMAP upper limits at
microwave frequencies. These upper limits were derived from the WMAP three-year data
(\cite{Hinshaw06}) considering that the fluctuation maps in all WMAP channels do not show
any excess at the location of the source and measuring the local noise in a nearby
circular region of 1 degree in diameter (De Bernardis, private communication). The
optical fluxes derived from the SDSS $z$, $i$ and $r$ magnitudes (the $g$ and $u$
measurements were not used since these are heavily affected by Lyman $\alpha$~ forest
absorption) are consistent with the UVOT upper limits (particularly with the $B$ value)
and describe a flat continuum with power that is lower than that at X-ray frequencies,
implying that the overall SED cannot peak near the optical band.

The radio through X-ray SED of \roxa ~shown in Fig.\ref{SED} can be described by a single
component, well approximated by a power-law, that bends only after the UV (dashed line)
and reaches the X-ray band with a photon index still rather flat. This is also supported
by the consideration that the emitted soft X-ray emission could be larger than that shown
in Fig.\ref{SED} because of the likely presence of intrinsic and intergalactic absorption
(see Sect. 2 and 3.2).

This simple description suggests that the emission is likely due to synchrotron radiation
with power that peaks at energies higher than 10 keV ($\sim 50$ keV in the rest-frame).
This is an extremely high energy that has been observed in rare cases during strong
flares only in nearby blazars with luminosity at least two orders of magnitudes lower
that that of \roxa~ (e.g. MKN501, \cite{Pian98,Mas04b}).

The sparse data coverage of the SED, however, leaves room to alternative scenarios for
the origin of the broad band emission of this source: in particular, one cannot exclude
that X-ray emission is due inverse Compton scattering either in a SSC or in external
Compton scenarios. In these cases one has to consider the possibility that the peak of
the synchrotron emission would be in the far infra-red with power of the order of $\sim
10^{-12}$ \ergs (see the dotted curve in Fig. \ref{SED}) where no measurements are
available, nor useful upper limits can be derived from existing data. However, we regard
it as unlikely since in this case : i) the optical spectrum would be in the steeply
declining part of the synchrotron component with a slope $\alpha_o$ larger than 2, which
is only marginally consistent with Fig. \ref{sdss_spectrum}  (see also Fig. \ref{SED})
even without de-reddening the spectrum (see Sect. 2); ii) the \fxfr~ ratio of \roxa ~is
much larger than that of known blazars with inverse Compton X-ray spectrum (e.g.,
\cite{gio06c,Gio06a}); iii) in Fig. \ref{SED} the boxes drawn at microwave frequencies
represent the flux expected from the observed ($\pm 1 \sigma$: solid box and
full-observed-range: dotted box) X-ray to microwave flux ratios in a large sample of WMAP
detected sources (\cite{gio06c,Gio06a}) when the X-rays are due to inverse Compton
emission; iv) in the hypothesis that the X-rays are Compton up-scattered microwave-far
infrared photons one expects the SED to be peaked in the far infrared. In this case, the
low energy spectrum of \roxa~ consistently with the WMAP upper limits, should
dramatically flatten after the radio (cm) band and become strongly inverted at microwave
frequencies (Fig.\ref{SED}).

The assumption that the X-ray emission from \roxa~ is not due to synchrotron radiation
implies that this object must be characterized by a peculiar overall SED with \fxfr ~and
\fxfmu orders of magnitude larger than in normal LBL blazars, and a highly inverted
microwave to far-infrared spectrum.  Therefore, it should be considered an unusual blazar
or possibly the prototype of a new class of radio loud AGN.

\begin{figure}[h]
 \begin{center}
 \epsfig{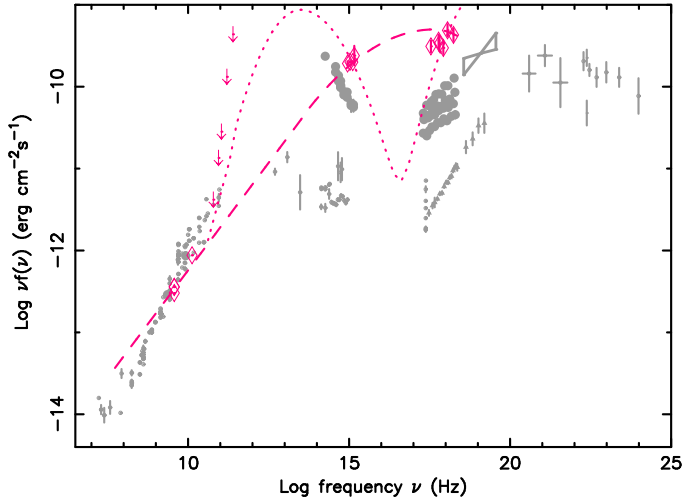}
 \end{center}
 \caption[]{The SED of \roxa~
(see Fig. \ref{SED}) is plotted together with that of 3C454.3 after scaling to match the
radio flux and the redshift of this last blazar. The large filled circles are the fluxes
observed during the very large flare of 3C454.3 in May 2005, while the smaller symbols
are non-simultaneous historical data taken from the literature (see \cite{gio06b} for
details). }
 \label{3c454_comp}
\end{figure}

A further possibility is to assume that \roxa~ was caught during an extremely large
optical and X-ray flare like that of 3C454.3 in May 2005 (\cite{pian06,gio06b}). To test
this hypothesis, we plotted in Fig. \ref{3c454_comp} the SED of both sources after
scaling the flux of \roxa~ to match the radio flux and redshift of 3C454.3. Although the
rescaled optical and X-ray fluxes are still substantially larger than that of 3C454.3, an
even larger flare cannot be excluded. However, we regard this explanation as unlikely
since both the optical and X-ray flux of 3C454.3 during the flare of 2005 varied by large
factors over time scales of hours or days, while \roxa~  was found approximately at the
same flux level in observations separated by several years both in the optical (the
source is clearly visible close to the POSS limiting sensitivity of  $R \simeq 20$ mag
and was detected as a $r=19.7$ mag object in the SDSS) and in the X-ray band by ROSAT and
Swift.

The most likely conclusion seems that \roxa ~is the first high-luminosity FSRQ with
synchrotron peak in the hard X-ray band. Its X-ray luminosity of $\sim 1.5\times 10^{47}$
erg/s, one of the largest ever observed in any AGN, has been found to be persistent and
approximately constant between the ROSAT and the \sw observation which are separated by
$\sim 10$ years.

\begin{figure}[h]
 \begin{center}
 \epsfig{file=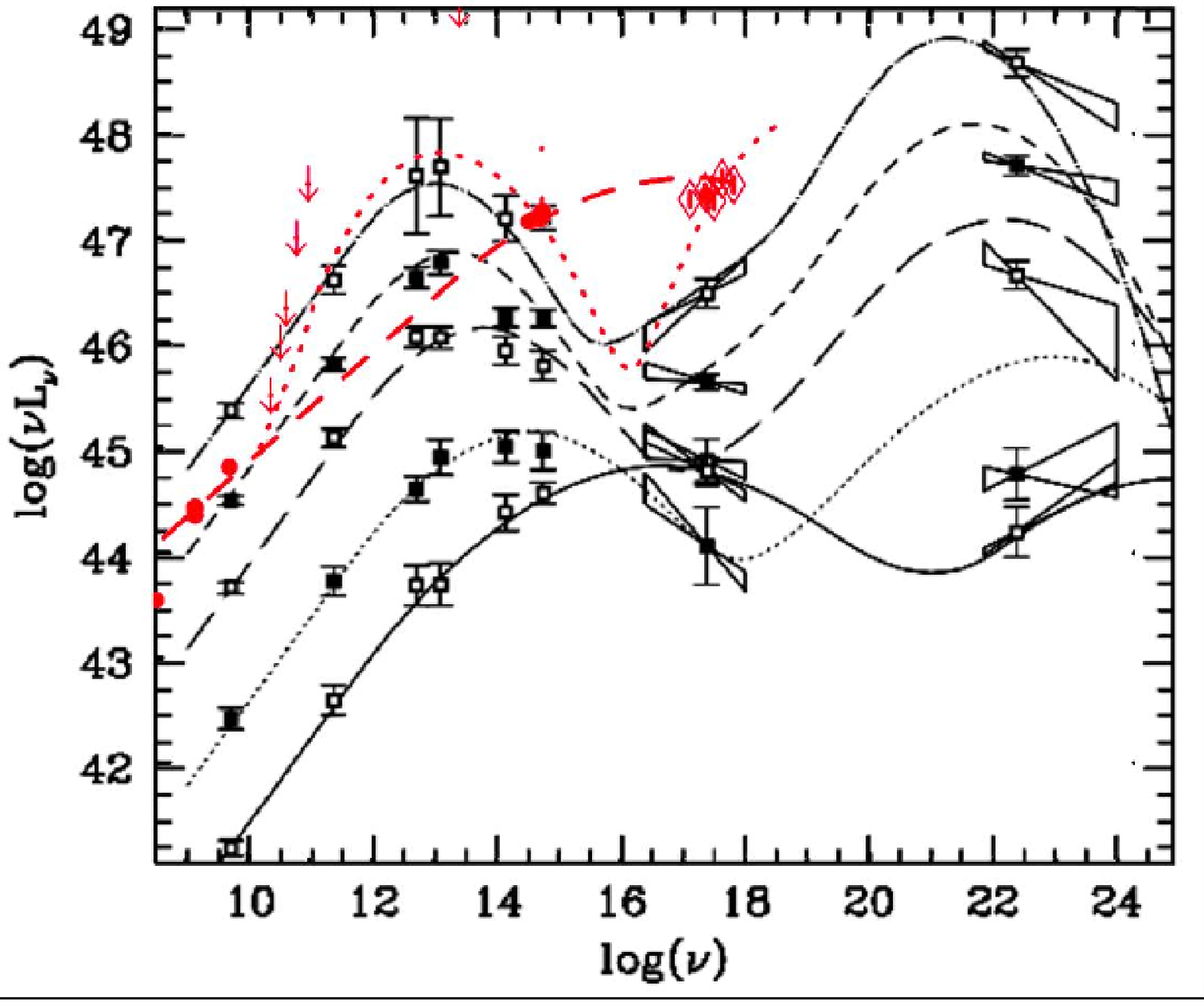,height=7.cm,angle=0}
 \end{center}
 \caption[]{The SED of \roxa~ shown Fig. \ref{SED}
 (converted to luminosity) is overlaid to the Blazar sequence
 plot of \cite{Fos98}. }
 \label{SED_bl_seq}
\end{figure}

An important consequence of the likely scenario where   the synchrotron peak of \roxa~ is
in the hard X-ray band is that  this object would strongly contradict the claimed
anti-correlation between luminosity and synchrotron peak energy, often referred to as the
"Blazar sequence" (\cite{Fos98,Ghi98}, see Fig.\ref{SED_bl_seq}). We recall that the
validity of this scenario has also been questioned by a number of studies involving large
samples of blazars selected in a variety of ways (e.g.,
\cite{Gio99,P03,Cac04,Nieppola05,Gio06a,Pad07}).

If the synchrotron emission of \roxa~ indeed peaks in the hard X-ray band,
then the peak of the inverse Compton component in a SSC scenario would
well be into the $\gamma$-ray band.
This object is therefore a very good target for GLAST to determine the
spectral shape of the inverse Compton component.

The possibility that such a high-energy synchrotron peak would be due
to a very high Doppler beaming, say $\delta > 50$, is not satisfactory
because the optical continuum would be so intense to overwhelm the
emission lines, which instead are very prominent.
If the beaming is instead more typical of blazar-like objects (e.g. $\delta
\simeq 10$), then \roxa~ should be capable of accelerating electrons
up to the TeV range and therefore it is similar to the near HBL sources
detected in this energy range.

The presently available data do not allow us to describe the SED of \roxa~ with a coverage
that is sufficient to safely establish its nature.
Future detailed (observational and theoretical) studies of such a peculiar
object will hence shed light on its energetic, relativistic effects and the
cosmological impact.

It is also important to know whether it is a unique source or the prototype of new class
of very high luminosty AGNs. The search for other high--$z$ FSRQ, based on a
multifrequency approach, is therefore of crucial relevance for the physics of this family
of extragalactic blazar-like sources.

\begin{acknowledgements}

The authors acknowledge the financial support for ASDC by the Italian Space Agency (ASI).
We are grateful to  P. De Bernardis for providing the WMAP upper limits. We thank G.
Fossati for providing an electronic version of the "Blazar sequence" figure originally
published in \cite{Fos98}. This work is partly based on data taken from the NVSS, FIRST,
ROSAT and SDSS archives. We thank the anonymous Referee for useful comments and
suggestions.
\end{acknowledgements}

\end{document}